\documentclass[lettersize,journal]{IEEEtran}
\usepackage{amsmath,amsfonts}
\usepackage{algorithmic}
\usepackage{algorithm}
\usepackage{array}
\usepackage[caption=false,font=normalsize,labelfont=sf,textfont=sf]{subfig}
\usepackage{textcomp}
\usepackage{stfloats}
\usepackage{url}
\usepackage{verbatim}
\usepackage{graphicx}
\usepackage{cite}

\usepackage[table]{xcolor}
\usepackage{multirow}

\begin{document}

\title{Reflecthernet: Exfiltrating 100BASE-TX Ethernet Traffic Using a Retroreflector Hardware Trojan} 

\author{\IEEEauthorblockN{Pierre Granier, Matthieu Davy, Philippe Besnier, François Sarrazin}\\
\IEEEauthorblockA{Univ Rennes, INSA Rennes, CNRS, IETR - UMR 6164, F-35000 Rennes, France\\
Email: \{pierre.granier, matthieu.davy, francois.sarrazin\}@univ-rennes.fr philippe.besnier@insa-rennes.fr}
 
}

\markboth{This work has been submitted to the IEEE for possible publication. Copyright may be transferred without notice.}%
{Shell \MakeLowercase{\textit{et al.}}: A Sample Article Using IEEEtran.cls for IEEE Journals}


\maketitle

\begin{abstract}
\textit{This work has been submitted to the IEEE for possible publication. Copyright may be transferred without notice, after which this version may no longer be accessible.} Electromagnetic eavesdropping is a well-established attack vector for remotely monitoring a target activity, most notably displays, over considerable ranges.
Other targets have been considered resistant to such attacks or do not exhibit sufficient electromagnetic leakage for practical exploitation.
Radio-frequency retroreflector attacks (RFRA) were developed to enable covert, active monitoring of a target by implanting a minimal hardware Trojan. 
These implants, typically implemented using discrete components such as transistors or diodes, do not betray their presence by emitting signals themselves; rather, they modulate the electromagnetic reflectivity of the target depending on the probed signal line data. Prior RFRA work has demonstrated their viability against video links and low-speed peripheral interfaces. In this work, we extend the applicability of RFRA to high-speed targets by presenting a successful attack on the 100BASE-TX Ethernet  standard.
We describe the design and realization of a compact implant capable of recovering the MLT-3 encoded signaling used in Fast Ethernet, as well as a dedicated demodulation and interpretation pipeline that mitigates errors introduced by the radio channel and maximizes the amount of recovered information. Experimental results validate the feasibility of covertly monitoring Fast Ethernet traffic using RF retroreflection and highlight the viability of such attacks for high-speed links.
\end{abstract}


\section{Introduction}

Remote eavesdropping of a device's confidential information through its unintended electromagnetic emanations is a long-standing concern in information security. Known broadly as TEMPEST attacks, these techniques exploit the fact that electronic components (cables, connectors and circuit traces) radiate signals that correlate with the processed data. An adversary equipped with suitable antennas and receivers can capture these radiated signals from a distance and, using signal-processing and reconstruction techniques, recover sensitive information.

Display devices have long been identified as significant sources of compromising electromagnetic emanations. The seminal work of Wim van Eck in 1985 \cite{VanEck1985} demonstrated that images rendered on CRT monitors could be reconstructed from their radiated signals. Subsequent studies revealed that both analog interfaces (e.g., VGA) and digital standards (e.g., DVI, HDMI, LVDS) remain susceptible to such leakage \cite{Kuhn2004ElectromagneticER, kuhn_compromising_2011, Lee2018} despite the adoption of high-speed differential signaling. More recently, it was shown that even scrambled interfaces such as DisplayPort can emit discernible patterns correlated with displayed content \cite{IEEEexample:erdeljan2023,erdeljan_electromagnetic_2026}, indicating that electromagnetic eavesdropping persists as a practical threat across successive generations of display technology. While such attacks have been shown to be practical on video links, compromising emanations are not limited to display interfaces. Prior work has demonstrated similar vulnerabilities in a variety of communication channels, including serial links such as RS-232 \cite{smulders_threat_1990}, keyboard interfaces like PS/2 and USB \cite{Vuagnoux2009}, and even low-speed network connections, where 10BASE-T Ethernet transmissions (10 Mbit/s) have been shown to leak recoverable data through their electromagnetic signatures \cite{10.1145/2939918.2940650}.

While TEMPEST attacks exploit a device's unintentional electromagnetic emanations, they are inherently limited: the signals are typically weak, noisy and dependent on the target shielding, making data recovery challenging. Indeed, an attacker can only passively observe the device emanations, with no ability to amplify or encode the information, and success often depends on proximity, sensitive equipment, and favorable environmental conditions. These constraints limit the practical applicability of passive eavesdropping, particularly for low-power or shielded devices. In contrast, attacks based on a hardware Trojan (HT), i.e., a malicious modification of a circuit, provides a more controlled and deliberate means of exfiltration.
The most straightforward implementation of such an implant would consist of one or more integrated circuits (ICs) capable of parsing the targeted link and transmitting relevant data over a radio link. However, this approach requires a powered implant, a non-negligible physical footprint and an active radio link, making it easily detectable.
 
Retroreflector implants are a subset of HTs that, contrary to the previously-described structure, do not directly transmit anything, do not require a power rail, and present a smaller footprint. The principle was first demonstrated with the infamous Great Seal Bug \cite{nikitin_leon_2012}, which consisted of an acoustic cavity whose properties were modulated by ambient sound via a thin moving diaphragm. The cavity was positioned at the end of a wire antenna, effectively acting as a variable impedance. By illuminating the device and recording the backscattered signal, an attacker could remotely eavesdrop on conversations within the bugged room. Retroreflector HTs gained wider attention through the leaked NSA ANT catalog documents \cite{IEEEexample:NSA}, which described retroreflector HT implementations capable of recovering low data-rate information, including approximate device positioning, audio, and video content from VGA interfaces. These attacks have been referred to as Radio Frequency Retroreflector Attacks (RFRAs). In the academic literature, RFRAs have been demonstrated on different technologies, including USB interfaces \cite{Wakabayashi_usenix_rfra_2018}, VGA displays \cite{IEEEexample:NSA, Kinugawa_Fujimoto_Hayashi_2019,11176387}, and audio systems \cite{Kinugawa_Fujimoto_Hayashi_2019, wang_wireless_2025}. Therefore, they have been limited to either analog targets (in particular VGA, which can exhibit relatively high data rates but where the attack success does not require pixel-level recovery) or digital links with a maximum symbol rate of a few MBd (see Table~\ref{tab:prev_rate}). Also, all targeted digital links were limited to a simple binary encoding scheme. 

\textbf{Motivations:} The primary motivation behind this work is to extend the attack surface of RFRA to high-speed links and more complex physical-layer encodings, by targeting high-speed Ethernet links. Indeed, Ethernet \cite{9844436} remains one of the most widely deployed standards, connecting desktops, servers, industrial controllers, IoT devices, and critical infrastructures \cite{ZUNINO2020103433}. Although modern Ethernet deployments increasingly employ encryption protocols such as TLS, IPsec, or MACsec to protect sensitive traffic, a portion of Ethernet communication remains unencrypted. In particular, wired internal networks (either partially isolated or even air-gapped) are sometimes considered “safe” and therefore continue to rely on legacy protocols and applications, either by necessity or oversight. To our knowledge, no large-scale studies have systematically quantified the volume of encrypted traffic in enterprise, industrial, or embedded networks; however, it is reasonable to assume that coverage is uneven. Indeed, within enterprise networks, unencrypted DNS traffic is expected, as encrypted DNS adoption is still not widespread even outside private environments \cite{Lyu_2022}. Legacy unencrypted protocols such as NFSv2/3, HTTP, or misconfigured SMB traffic can still be found in enterprise networks. Depending on the layer at which encryption is used, unencrypted lower layers can enable passive packet fingerprinting, leaking information useful for an attacker \cite{anderson2017osfingerprintingnewtechniques}. The combination of widespread deployment and persistent use in legacy and embedded systems motivates a focused investigation into the eavesdropping capabilities against Ethernet traffic.

\textbf{Attack model and scenarios:} 
\begin{itemize}
\item \textit{Targeted links:} 
The proposed implant architecture is designed to be embedded within one of the two differential pairs used in a 100BASE-TX Ethernet link, enabling the recovery of traffic in one direction. 
In the case of deployment on a 1000BASE-T link, additional hardware modifications can be introduced to force a downgrade to the vulnerable 100BASE-TX standard, thereby allowing the implant to operate on most of the commonly deployed Ethernet links. 
\item \textit{Implementation scenarios:}  
The proposed hardware Trojan (HT) may be integrated at the cable manufacturing or distribution, corresponding to a supply chain attack.
Alternatively, the implant may be introduced when temporary access to the target is possible, enabling modification or replacement of an existing link (i.e., an evil maid attack). 
\item \textit{Attack setup:}  
Following implantation, no further physical access is required, and the implant remains inert unless interrogated. 
Interrogation is performed using two antennas and a high-end software-defined radio (SDR) capable of phase-coherent transmission and reception, operating at an appropriate sampling rate.
\end{itemize} 

\textbf{Contributions} In this work, we demonstrate the practicality of targeting 100BASE-TX Ethernet links using a specially-designed retroreflector HT. This implant translates signals from one end of the Ethernet link into a modulation of its host's backscattering behavior, enabling the recovery of transmitted data through an RFRA \cite{Wakabayashi_usenix_rfra_2018}. This paper makes the following key contributions:

\begin{itemize}
\item \textit{First retroreflector attack on Ethernet:} We present a complete framework, called Reflecthernet, capable of exfiltrating 100BASE-TX Ethernet traffic using RFRAs. By successfully exfiltrating data at a 125~MBd symbol rate, we extend the applicability of RFRA to data rates an order of magnitude higher than those previously reported for RFRA targeting digital links (see Table~\ref{tab:prev_rate}).
\item \textit{Attack range evaluation:} We demonstrate that Reflecthernet can achieve near-complete data recovery at a range of 3~meters in an office environment with limited transmitted power. Extensive measurements in a shielded anechoic chamber show that the attack range can be significantly extended by increasing the injected power.
\end{itemize}

In order to achieve this, Reflecthernet differs from prior work in three key aspects:
\begin{itemize}
\item \textit{Novel HT architecture:} We design and implement a new HT architecture based on a simple two-diode structure, enabling effective discrimination and modulation of MLT-3 signaling with minimal complexity.
\item \textit{High-bandwidth interrogation setup:} We use an attack platform built around an RF System-on-Chip (RFSoC) evaluation board (ZCU111), which supports sampling rates of up to multiple~GSa/s. This broadens the range of protocols that can be targeted by RFRAs.
\item \textit{Protocol-aware error correction:} We propose a custom demodulation and decoding pipeline that includes an error-correcting mechanism and an error-resistant scrambler-state reconstruction, enabling reliable recovery of the transmitted bit stream by mitigating the impact of ambient noise and signal distortions on the backscatter channel.
\end{itemize}

\begin{table}[h]
\begin{tabular}{llllll}
\hline
\multicolumn{1}{|c|}{\textbf{\begin{tabular}[c]{@{}c@{}}Link \\ targeted\end{tabular}}}         & \multicolumn{1}{l|}{\textbf{\begin{tabular}[c]{@{}l@{}}Analog /\\ Digital\end{tabular}}}  & \multicolumn{1}{c|}{\textbf{\begin{tabular}[c]{@{}c@{}}Targeted \\ Symbol Rate\end{tabular}}}     & \multicolumn{1}{c|}{\textbf{Work}}                     & \multicolumn{1}{l|}{\textbf{\begin{tabular}[c]{@{}l@{}}Attack\\Sample Rate\end{tabular}}}  \\ \hline
\multicolumn{1}{|l|}{\multirow{4}{*}{VGA}}           & \multicolumn{1}{l|}{\multirow{4}{*}{Analog}}  & \multicolumn{1}{l|}{?}                                 & \multicolumn{1}{l|}{\cite{IEEEexample:NSA}}                   & \multicolumn{1}{l|}{?}  \\ \cline{3-5}
\multicolumn{1}{|l|}{}                               & \multicolumn{1}{l|}{}                         & \multicolumn{1}{l|}{?}                                 & \multicolumn{1}{l|}{\cite{OSS_RFRA}}                         & \multicolumn{1}{l|}{20~Msp/s}                \\ \cline{3-5} 
\multicolumn{1}{|l|}{}                               & \multicolumn{1}{l|}{}                         & \multicolumn{1}{l|}{$\sim$47~MBd}             & \multicolumn{1}{l|}{\cite{Kinugawa_Fujimoto_Hayashi_2019}} & \multicolumn{1}{l|}{\textgreater 100~Msp/s}  \\ \cline{3-5} 
\multicolumn{1}{|l|}{}                               & \multicolumn{1}{l|}{}                         & \multicolumn{1}{l|}{$\sim$38~MBd}             & \multicolumn{1}{l|}{\cite{11176387}}                          & \multicolumn{1}{l|}{40~Msp/s}  \\ \hline
\multicolumn{1}{|l|}{\begin{tabular}[c]{@{}l@{}}Audio\\ (PDM signal)\end{tabular}}             & \multicolumn{1}{l|}{Digital}                  & \multicolumn{1}{l|}{2~MBd}                          & \multicolumn{1}{l|}{\cite{Kinugawa_Fujimoto_Hayashi_2019}} & \multicolumn{1}{l|}{10~Msp/s}\\ \hline
\multicolumn{1}{|l|}{\multirow{3}{*}{\begin{tabular}[c]{@{}l@{}}USB \\ Low-Speed\end{tabular}}} & \multicolumn{1}{l|}{\multirow{3}{*}{Digital}} & \multicolumn{1}{l|}{\multirow{3}{*}{1.5~MBd}} & \multicolumn{1}{l|}{\cite{OSS_RFRA}}                         & \multicolumn{1}{l|}{20~Msp/s}          \\ \cline{4-5} 
\multicolumn{1}{|l|}{}                               & \multicolumn{1}{l|}{}                         & \multicolumn{1}{l|}{}                                  & \multicolumn{1}{l|}{\cite{Wakabayashi_usenix_rfra_2018}}   & \multicolumn{1}{l|}{10~Msp/s}          \\ \cline{4-5} 
\multicolumn{1}{|l|}{}                               & \multicolumn{1}{l|}{}                         & \multicolumn{1}{l|}{}                                  & \multicolumn{1}{l|}{\cite{Kinugawa_Fujimoto_Hayashi_2019}} & \multicolumn{1}{l|}{50~Msp/s}    \\ \hline
\multicolumn{1}{|l|}{PS/2}                           & \multicolumn{1}{l|}{Digital}                  & \multicolumn{1}{l|}{16.7~kBd}                 & \multicolumn{1}{l|}{\cite{OSS_RFRA}}                         & \multicolumn{1}{l|}{20~Msp/s}    \\ \hline
\multicolumn{1}{|l|}{100BASE-TX}                     & \multicolumn{1}{l|}{Digital}                  & \multicolumn{1}{l|}{125~MBd}                  & \multicolumn{1}{l|}{\begin{tabular}[c]{@{}l@{}}This \\ work\end{tabular}}                         & \multicolumn{1}{l|}{1.6~Gsp/s}   \\ \hline
                                                     &                                               &                                                        &                                                        &                                               
\end{tabular}
\caption{Summary of previous RFRA attacks and targets, detailing the targeted link characteristics and used sample rate per attack.
For VGA, the pixel rate can vary widely, hence, different targeted symbol rate per attack.}
\label{tab:prev_rate}
\end{table} 

\textbf{Outline:} The remainder of this paper is organized as follows. Section~\ref{sec:background} presents the necessary background, including the operating principles of retroreflector HTs and a detailed overview of the 100BASE-TX encoding and decoding schemes. The subsequent sections describe the main technical contributions of this work: the design of the novel HT architecture (Section~\ref{sec:trojan_archi}), the high-speed interrogation setup (Section~\ref{sec:implant_inter}), the data recovery methodology (Section~\ref{sec:data_recovery}) and the application of the attack on an active link in an office environment (Section~\ref{sec:realistic}). Finally, Section~\ref{sec:conclusion} concludes the paper and discusses potential future directions.

\section{Background}
\label{sec:background}

\subsection{RF Retroreflector Attack}

A retroreflector HT functions as a backscattering element, analogous to a UHF RFID tag. It consists of an antenna with impedance $Z_\mathrm{ant}$ loaded by an impedance $Z_\mathrm{load}$. The corresponding reflection coefficient $\Gamma$ is given by
\begin{equation}
    \Gamma = \frac{Z_{\mathrm{load}} - Z^{*}_{\mathrm{ant}}}{Z_{\mathrm{load}} + Z_{\mathrm{ant}}}.
\end{equation}

\begin{figure*}[!t]
\centering
\includegraphics[width=\textwidth]{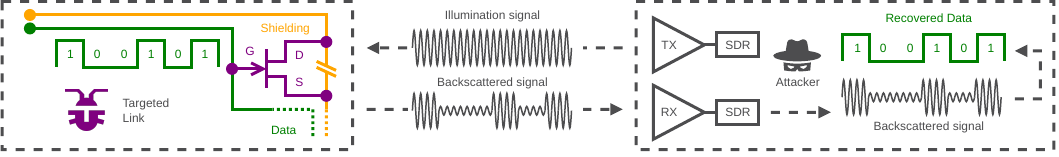}
\caption{Schematic of an RFRA in which a transistor-based HT allows monitoring of secret data through a backscattered signal.
The implant modulates the backscattering characteristics of a shielding section by switching state according to the logic level applied to the gate by the secret data, switching the impedance between drain and source from low to high.
Sending the illumination signal and receiving the echo is done with the help of two separate antennas connected to one or multiple SDRs.}
\label{fig:principle}   
\end{figure*}

If the antenna and load impedances are matched ($Z_\mathrm{load} = Z^{*}_\mathrm{ant}$), then $\Gamma = 0$, indicating that no reflection occurs. Conversely, when $Z_\mathrm{load} \to \infty$ (open-circuit condition), $\Gamma = 1$, and the backscattering element completely reflects the incident signal. 
Therefore, a retroreflector HT requires a variable load whose impedance changes according to the secret data to be exfiltrated. In the literature, this is typically achieved by employing a transistor (typically a FET transistor) where the drain and source terminals are connected to an improvised antenna (usually made from the host shielding \cite{Kinugawa_Fujimoto_Hayashi_2019,IEEEexample:NSA}). The transistor's conduction state is controlled by the compromised signal line connected to its gate terminal, as illustrated in Fig~\ref{fig:principle}.

When the data line carries a logical  '0', the transistor is in the OFF state and presents a drain–source impedance with its corresponding reflection coefficient $\Gamma_{\mathrm{off}}$. Conversely, when the line carries a logical  '1', the transistor switches ON, resulting in a different reflection coefficient $\Gamma_{\mathrm{on}}$. The variation between these two impedance states directly mirror the digital information transmitted on the data line. When the circuit is illuminated by an external RF signal to perform an RFRA, as illustrated in Fig~\ref{fig:principle}, this impedance modulation produces a corresponding modulation in the backscattered wave. As a result, the confidential data can be reconstructed by demodulating the backscattered signal.

Two practical points must be emphasized. First, although the RFRA mechanism is presented here primarily as an amplitude modulation process, the reflection coefficient is in fact complex, resulting in both amplitude and phase modulation components. In practice, the actual reflection behavior depends on the geometry and material properties of the improvised antenna (such as the cable shielding repurposed as a dipole) which are typically uncontrolled. Consequently, the reflection coefficients never correspond to the idealized cases of $\Gamma = 0$ or $\Gamma = 1$, leading to non-ideal but information-bearing backscatter modulation.
Second, the implant's complex response is overlapped on the environment's non-modulating response (direct TX→RX coupling, plus static reflections from surrounding objects). 
This static background offsets the implant response both in magnitude and phase, the implant's modulation appears then as a small, time-varying perturbation on top of a generally much larger, static term.
This implies that the absolute complex response measured at the receiver is unpredictable a priori even for a carefully designed implant, and that the large unmodulated background can dominate the receiver front end and cause saturation (i.e., “blinding” the receiver).

\subsection{100BASE-TX operation} 
\label{sec:ethernet_operation}

\subsubsection{Physical layer encoding}

The 100BASE-TX standard uses  Category 5 (or higher) Ethernet cables. Although such cables typically contain four twisted pairs, only two are actually used (one for each direction), enabling full-duplex operation.

\begin{figure}[!t]
\centering
\includegraphics[width=\columnwidth]{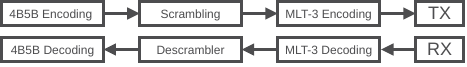}
\caption{Summary of the 100BASE-TX encoding and decoding steps.}
\label{fig:encode_decode}   
\end{figure} 

Any data originating from the Data Link layer and transmitted over a Ethernet 100BASE-TX link passes through dedicated hardware that prepares it for physical transmission. This process involves multiple encoding and signal transformation stages, as specified by IEEE Std 802.3 \cite{9844436} and ANSI INCITS 263-1995 (TP-PMD).
The transmission process comprises three main stages: a 4B/5B encoding step based on a predefined symbol dictionary, a scrambling operation implemented using a linear feedback shift register (LFSR), and finally an MLT-3 line encoding for physical signaling. These stages are summarized in Fig~\ref{fig:encode_decode} and detailed in the following subsections.

\paragraph{\textbf{4B/5B encoding}} 
 
\begin{table}[]
\begin{tabular}{clcc}
\hline
\multicolumn{4}{|c|}{\textbf{DATA}}                                                                                                                             \\ \hline
\multicolumn{1}{|c|}{5-bit codes} & \multicolumn{1}{l|}{Name} & \multicolumn{1}{c|}{4-bit nibbles} & \multicolumn{1}{c|}{Interpretation}                \\ \hline
\multicolumn{1}{|c|}{11110}      & \multicolumn{1}{l|}{0}    & \multicolumn{1}{c|}{0000}          & \multicolumn{1}{c|}{Data 0}                        \\ \hline
\multicolumn{1}{|c|}{01001}      & \multicolumn{1}{l|}{1}    & \multicolumn{1}{c|}{0001}          & \multicolumn{1}{c|}{Data 1}                        \\ \hline
\multicolumn{1}{|c|}{10100}      & \multicolumn{1}{l|}{2}    & \multicolumn{1}{c|}{0010}          & \multicolumn{1}{c|}{Data 2}                        \\ \hline
\multicolumn{1}{|c|}{10101}      & \multicolumn{1}{l|}{3}    & \multicolumn{1}{c|}{0011}          & \multicolumn{1}{c|}{Data 3}                        \\ \hline
\multicolumn{1}{|c|}{01010}      & \multicolumn{1}{l|}{4}    & \multicolumn{1}{c|}{0100}          & \multicolumn{1}{c|}{Data 4}                        \\ \hline
\multicolumn{1}{|c|}{01011}      & \multicolumn{1}{l|}{5}    & \multicolumn{1}{c|}{0101}          & \multicolumn{1}{c|}{Data 5}                        \\ \hline
\multicolumn{1}{|c|}{01110}      & \multicolumn{1}{l|}{6}    & \multicolumn{1}{c|}{0110}          & \multicolumn{1}{c|}{Data 6}                        \\ \hline
\multicolumn{1}{|c|}{01111}      & \multicolumn{1}{l|}{7}    & \multicolumn{1}{c|}{0111}          & \multicolumn{1}{c|}{Data 7}                        \\ \hline
\multicolumn{1}{|c|}{10010}      & \multicolumn{1}{l|}{8}    & \multicolumn{1}{c|}{1000}          & \multicolumn{1}{c|}{Data 8}                        \\ \hline
\multicolumn{1}{|c|}{10011}      & \multicolumn{1}{l|}{9}    & \multicolumn{1}{c|}{1001}          & \multicolumn{1}{c|}{Data 9}                        \\ \hline
\multicolumn{1}{|c|}{10110}      & \multicolumn{1}{l|}{A}    & \multicolumn{1}{c|}{1010}          & \multicolumn{1}{c|}{Data A}                        \\ \hline
\multicolumn{1}{|c|}{10111}      & \multicolumn{1}{l|}{B}    & \multicolumn{1}{c|}{1011}          & \multicolumn{1}{c|}{Data B}                        \\ \hline
\multicolumn{1}{|c|}{11010}      & \multicolumn{1}{l|}{C}    & \multicolumn{1}{c|}{1100}          & \multicolumn{1}{c|}{Data C}                        \\ \hline
\multicolumn{1}{|c|}{11011}      & \multicolumn{1}{l|}{D}    & \multicolumn{1}{c|}{1101}          & \multicolumn{1}{c|}{Data D}                        \\ \hline
\multicolumn{1}{|c|}{11100}      & \multicolumn{1}{l|}{E}    & \multicolumn{1}{c|}{1110}          & \multicolumn{1}{c|}{Data E}                        \\ \hline
\multicolumn{1}{|c|}{11101}      & \multicolumn{1}{l|}{F}    & \multicolumn{1}{c|}{1111}          & \multicolumn{1}{c|}{Data F}                        \\ \hline
\multicolumn{1}{|c|}{11111}      & \multicolumn{1}{l|}{I}    & \multicolumn{1}{c|}{Undefined}     & \multicolumn{1}{c|}{IDLE inter-stream fill code}   \\ \hline
\rowcolor{gray!30} \multicolumn{1}{|c|}{00000}      & \multicolumn{1}{l|}{P}    & \multicolumn{1}{c|}{0001}          & \multicolumn{1}{c|}{SLEEP}                         \\ \hline
                                 &                           &                                    &                                                    \\ \hline
\multicolumn{4}{|c|}{\textbf{CONTROL}}                                                                                                                          \\ \hline
\multicolumn{1}{|c|}{5bit codes} & \multicolumn{1}{l|}{Name} & \multicolumn{1}{c|}{4 bit nibbles} & \multicolumn{1}{c|}{Interpretation}                \\ \hline
\multicolumn{1}{|c|}{11000}      & \multicolumn{1}{l|}{J}    & \multicolumn{1}{c|}{0101}          & \multicolumn{1}{c|}{Start of stream delimiter - 1} \\ \hline
\multicolumn{1}{|c|}{10001}      & \multicolumn{1}{l|}{K}    & \multicolumn{1}{c|}{0101}          & \multicolumn{1}{c|}{Start of stream delimiter - 2} \\ \hline
\multicolumn{1}{|c|}{01101}      & \multicolumn{1}{l|}{T}    & \multicolumn{1}{c|}{undefined}     & \multicolumn{1}{c|}{End of stream delimiter - 1}   \\ \hline
\multicolumn{1}{|c|}{00111}      & \multicolumn{1}{l|}{R}    & \multicolumn{1}{c|}{undefined}     & \multicolumn{1}{c|}{End of stream delimiter - 2}   \\ \hline
                                 &                           &                                    &                                                    \\ \hline
\multicolumn{4}{|c|}{\textbf{INVALID}}                                                                                                                          \\ \hline
\multicolumn{1}{|c|}{5bit codes} & \multicolumn{1}{l|}{Name} & \multicolumn{1}{c|}{4 bit nibbles} & \multicolumn{1}{c|}{Interpretation}                \\ \hline
\rowcolor{gray!30} \multicolumn{1}{|c|}{00100}      & \multicolumn{1}{l|}{H}    & \multicolumn{1}{c|}{Undefined}     & \multicolumn{1}{c|}{Force Transmit Error}          \\ \hline
\rowcolor{gray!30} \multicolumn{1}{|c|}{00001}      & \multicolumn{1}{l|}{}     & \multicolumn{1}{c|}{Undefined}     & \multicolumn{1}{c|}{Invalid code}                  \\ \hline
\rowcolor{gray!30} \multicolumn{1}{|c|}{00010}      & \multicolumn{1}{l|}{}     & \multicolumn{1}{c|}{Undefined}     & \multicolumn{1}{c|}{Invalid code}                  \\ \hline
\rowcolor{gray!30} \multicolumn{1}{|c|}{00011}      & \multicolumn{1}{l|}{}     & \multicolumn{1}{c|}{Undefined}     & \multicolumn{1}{c|}{Invalid code}                  \\ \hline
\rowcolor{gray!30} \multicolumn{1}{|c|}{00101}      & \multicolumn{1}{l|}{}     & \multicolumn{1}{c|}{Undefined}     & \multicolumn{1}{c|}{Invalid code}                  \\ \hline
\rowcolor{gray!30} \multicolumn{1}{|c|}{00110}      & \multicolumn{1}{l|}{}     & \multicolumn{1}{c|}{Undefined}     & \multicolumn{1}{c|}{Invalid code}                  \\ \hline
\rowcolor{gray!30} \multicolumn{1}{|c|}{01000}      & \multicolumn{1}{l|}{}     & \multicolumn{1}{c|}{Undefined}     & \multicolumn{1}{c|}{Invalid code}                  \\ \hline
\rowcolor{gray!30} \multicolumn{1}{|c|}{01100}      & \multicolumn{1}{l|}{}     & \multicolumn{1}{c|}{Undefined}     & \multicolumn{1}{c|}{Invalid code}                  \\ \hline
\rowcolor{gray!30} \multicolumn{1}{|c|}{10000}      & \multicolumn{1}{l|}{}     & \multicolumn{1}{c|}{Undefined}     & \multicolumn{1}{c|}{Invalid code}                  \\ \hline
\rowcolor{gray!30} \multicolumn{1}{|c|}{11001}      & \multicolumn{1}{l|}{}     & \multicolumn{1}{c|}{Undefined}     & \multicolumn{1}{c|}{Invalid code}                  \\ \hline
                                 &                           &                                    &                                                    \\ 
\end{tabular}
\caption{4B/5B translation table. Greyed rows are codes considered as errors in our custom decoding pipeline.}
\label{tab:4B5B_table}
\end{table} 

Regardless of the context, all data bytes are divided into 4-bit nibbles and encoded into 5-bit symbols according to Table~\ref{tab:4B5B_table}.
Due to this encoding scheme, each twisted pair transmits symbols at a rate of $125~\mathrm{MBd}$, resulting in an effective data rate of $100~\mathrm{Mbit/s}$.
 
In the targeted implementation, when bytes are serialized from the data stream (e.g., MAC layer packets) the resulting 4-bit nibbles are transmitted by sending the low-order nibble first, requiring reordering during reconstruction.   

When no packets are transmitted and the link remains in the IDLE state, a continuous sequence of 'I' symbols is sent. When a packet transmission begins, a preamble byte composed of the 'J' and 'K' nibbles is transmitted, followed by the payload data. At the end of the packet, a postamble consisting of the 'T' and 'R' symbols marks the transition to the IDLE state. The interval between frame delimiters, including the delimiters themselves, spans 96 bits and defines the minimum inter-packet gap (IPG), as specified in IEEE 802.3, Section 4.4.2 \cite{9844436}.  

\paragraph{\textbf{Scrambling}} 

\begin{figure}[!t]
\centering
\includegraphics[width=\columnwidth]{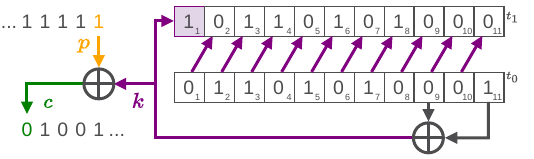}
\caption{
An example of the scrambler operation for a single bit. 
The scrambler output $k[0]$ at time $t[0]$ is computed from the bit 9 and 11 of it's internal state. 
The plain text is scrambled resulting in $p[0] \oplus k[0] = c[0]$. 
The scrambler internal state is then updated to time $t[1]$ shifting it's internal state and inserting $k[0]$.}
\label{fig:scrambler}   
\end{figure} 

After the 4B/5B translation, the bit stream is scrambled.
The purpose of the scrambler is not to provide security; rather, it helps create a pseudo-random distribution of logic levels to improve electromagnetic compatibility (EMC) by minimizing long sequences of identical bits (e.g., when the link is IDLE).
The scrambler is implemented using an 11-bit linear feedback shift register (LFSR) defined by the polynomial $x^{11} - x^{9} - 1$. 
At each clock cycle, the XOR of taps 9 and 11 is used as the scrambling bit and fed back into the LFSR input. This configuration produces a maximal-length pseudo-random sequence of  
$2^{11} - 1 = 2047$ bits before repeating. 

The scrambler input/output can be defined using:

\begin{itemize}
  \item $p[n]$: the plaintext bit sequence
  \item $k[n]$: the bit stream at the scrambler output
  \item $c[n]$: the scrambled bit-stream
\end{itemize}

The scrambling process, illustrated in Fig~\ref{fig:scrambler}, operates as follows:
Each plaintext bit $p[n]$ is XORed with the scrambler output $k[n]$, where  $k[n]  = k[n-9] \oplus k[n-11]$, resulting in the scrambled bit $c[n] = k[n] \oplus p[n]$. 
Each output bit updates the internal state of the LFSR, advancing the pseudo-random sequence.

\paragraph{\textbf{MLT-3}}  

The bit stream at the output of the scrambler is then converted into electrical signaling levels using a Multi-Level Transmit-3 (MLT-3) encoding scheme. In this scheme, data is represented by three discrete voltage levels, denoted as –1 ($D^{+} < D^{-}$), 0 ($D^{+} = D^{-}$), and +1 ($D^{+} > D^{-}$). 
The differential pair cycles through the following sequence of four states:
\[D^{+} > D^{-} \rightarrow D^{+} = D^{-} \rightarrow D^{+} < D^{-} \rightarrow D^{+} = D^{-}\]

The progression through this sequence depends on the transmitted data bit:
\begin{itemize}
  \item When transmitting a 0, the output level remains unchanged (the sequence “stalls”).
  \item When transmitting a 1, the output advances to the next state in the sequence.
\end{itemize}

This encoding scheme produces a pseudo-sine wave with a maximum frequency of: 
\[
f_\text{max} = \frac{125~\text{MHz (symbol rate)}}{4~\text{(shortest cycle in the sequence)}} = 31.25~\text{MHz}.
\]
  
\subsubsection{Physical layer decoding}
\label{sec:pld}

On the receiver side, all operations are applied in the reverse order to recover the original bit sequence. 

When decoding the MLT-3 logical levels, each transition within a bit period is interpreted as 1, while a constant logic level is interpreted as 0. 
Discriminating between –1 ($D^{+} < D^{-}$) and +1 ($D^{+} > D^{-}$) is unnecessary, as the intermediate state 0 ($D^{+} = D^{-}$) always occurs between them, preventing direct transitions from –1 to +1 or vice versa. 

The scrambler sequence must be recovered to decode each received bit.
The internal state of the scrambler can be inferred from known patterns in the transmitted data. In particular, during the mandatory IPG, a repeating IDLE pattern (all bits at one) is transmitted; this known sequence enables synchronization of the scrambler state using $k[n] = c[n] \oplus 1 \text{(IDLE)}$. 
The 96-bit IPG ensures that if scrambler synchronization is lost, it can be recovered after any subsequent packet transmission.  
Once the scrambler is synchronized, the original plaintext bits can be recovered using the relation $p[n] = k[n] \oplus c[n]$.

The 4B/5B translation table (Table~\ref{tab:4B5B_table}) is applied to all 5-bit codes. 
The alignment of the 5-bit sequences is required and performed each time the unscrambling operation yields a 0 (i.e., when a start of stream delimiter marks the end of the IDLE state).

\section{Implant architecture targeting MLT-3}
\label{sec:trojan_archi}

\begin{figure*}[!t]
\centering
\includegraphics[width=\textwidth]{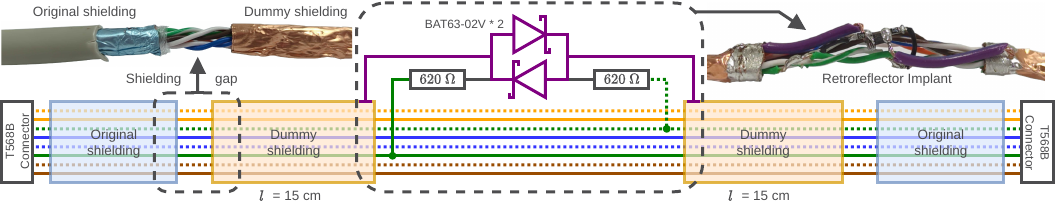}
\caption{A retroreflector implant composed of two Schottky diodes (BAT63-02V) and two $620~\Omega$ resistors. The implant is connected to an improvised antenna made from copper tape, which replaces the original cable shielding and is electrically isolated from it. The two antenna segments create a dipole of total length $2l~\mathrm{cm}$.}
\label{fig:implant}   
\end{figure*} 

Because of the three logic levels in the MLT-3 encoding, any implant architecture used for other targets \cite{IEEEexample:NSA,Kinugawa_Fujimoto_Hayashi_2019,Wakabayashi_usenix_rfra_2018,11176387,OSS_RFRA} would fail to accurately translate the logic levels without losing information. 
To overcome this, we developed the architecture shown in Fig~\ref{fig:implant}. Two Schottky diodes (BAT63-02V) of opposite polarity are connected to both lines of a differential MLT-3 pair, in this configuration either diode becomes conducting when $D^{+} \neq D^{-}$, translating the absolute differential voltage, $|D^{+} - D^{-}|$, into impedance change for $Z_\mathrm{load}$.  
Directly placing the diodes across the pair would create a short circuit, so a series resistance is required to preserve link operation; experimentally, two $620~\Omega$ resistors (one at each terminal of the variable load) provided sufficient isolation while still biasing the diodes. 
For very long cables, where signaling amplitude is naturally attenuated, the implant insertion may render the link non-functional; we did not investigate this effect on the maximum usable cable length.

The implant is installed by connecting a variable load into the differential pair: two wires connect the load to $D^{+}$ and $D^{-}$, while two additional conductors connect it to the antenna parts.
According to the cable construction, two types of improvised dipole antennas can be created.
 
For cables with an overall shield (F/*TP, S/*TP, SF/*TP) the original shielding may be disconnected at the implant location $\pm~l~\text{cm}$ or substituted to create an antenna of arbitrary composition. 
For unshielded cables (U/UTP), conductors from unused twisted pairs may be employed as the antenna; this approach has the side effect of preventing the negotiation of a 1000BASE-T link between the connected terminals. 

Intentionally preventing 1000BASE-T negotiation by disabling unused pairs help make the implant more context independent (allowing monitoring of original 100BASE-TX links and automatically downgrading 1000BASE-T) but this downgrade could be detected via software at both ends of the link, significantly decreasing covertness of the attack.

An example implant is shown in Fig~\ref{fig:implant}. In that implementation, we used an F/UTP cable whose shield was replaced by copper tape over a length $l$ = 15~cm, this implant design is at the basis for all results reported below.
The component used  in our implant is made from two BAT63-02V in an SC79 package (dimension 2~mm x 0.9~mm x 1.25~mm) and two resistors in a 0603 package (dimension 1.55~mm x 0.85~mm x 0.45~mm). These component sizes permit an implant to be embedded within Cat-5 (or higher) cable with minimal visible alterations.

\section{Implant interrogation}
\label{sec:implant_inter}

\begin{figure}[!t]
\centering
\includegraphics[width=\columnwidth]{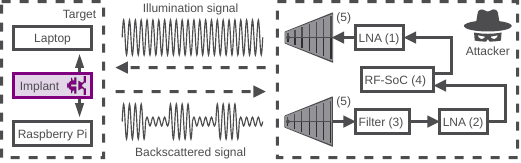}
\caption{Interrogation setup.
(1) OPA\_WBLNA wide-band LNA, $\sim 17$~dB gain on the transmitting path.
(2) PE15A63012 Broadband LNA, $\sim 40$~dB gain in the receiving path.
(3) SHP-300+ Lumped LC High Pass Filter, 290 - 3000 MHz.
(4) Illumination signal via a ZCU111 RFSoC evaluation board.
(5) LPDAMAX RF Space antennas (4-7 dBi over 300-1000 MHz).} 
\label{fig:interogation}  
\end{figure} 

\begin{figure}[!t]
\centering
\includegraphics[width=\columnwidth]{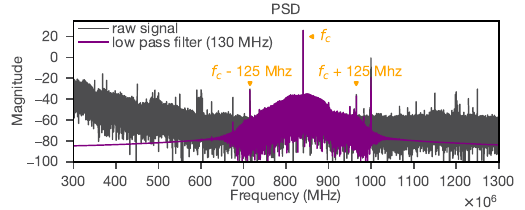}
\caption{Power spectral density (PSD) of the raw and filtered samples. 
This recording was made at a sample rate of $1.6~\mathrm{Gsp/s}$, with a center frequency $f_{c}$ of 840~MHz.
Significant amplitude is present at $f_{c} \pm$ 125~MHz due to the symbol rate of the 100BASE-TX link.}
\label{fig:PSD}   
\end{figure} 

Although MLT-3 signaling occupies a 31.25~MHz bandwidth, our implant architecture reduces the three-level encoding to a binary form by taking the absolute value $|D^{+} - D^{-}|$.
This operation yields a waveform with a maximum frequency of 125~MHz. 
Consequently, interrogation of the implant requires a receiver bandwidth of at least 125~MHz, and (to satisfy the Nyquist–Shannon sampling theorem) a sampling rate of at least 
$2 \times 125~\mathrm{MHz} = 250~\mathrm{Msp/s}$. 
In practice, higher sampling rates simplify signal processing and help yield better results.

These interrogation requirements exceed the capabilities of most SDRs, necessitating the use of high-performance hardware. In our implementation, we employed an RFSoC evaluation board (ZCU111), which supports sampling rates in the gigasample-per-second range.
To interface with this platform, the Xilinx 'RF Data Converter Evaluation Tool' was controlled via a custom Python API. This API enables bidirectional data exchange between the instrumenting computer and two synchronized ADC/DAC channels, allowing phase coherent interrogation.
Furthermore, because the transmitted waveform samples used for illumination are directly controlled, the illumination power can be dynamically adjusted by modifying the output waveform amplitude.
 
This setup is complemented by two low-noise amplifiers (LNAs), one on the transmit path ($\sim17$~dB gain) and one on the receive path ($\sim40$~dB gain). 
This configuration enables transmission at a carrier power of approximately 17~dBm over frequencies ranging from 300~MHz to 1~GHz.
On the reception side, the signal is filtered using an SHP-300+ high-pass filter, which attenuates frequencies below 300 MHz to suppress parasitic components that could otherwise saturate the receiving LNA. 
Both illumination and reception rely on wideband PCB log-periodic antennas.
The complete interrogation setup is illustrated in Fig~\ref{fig:interogation}.
 
Using this setup, illumination and data acquisition were performed at a sampling rate of 1.6~Gsp/s. 
This rate, although higher than strictly necessary, was selected to maximize the quality of the received data.
However, streaming and storing I/Q samples at this rate is technically demanding, we therefore rely solely on the limited DDR memory assigned to this ADC (128~MB in the default RF Data Converter Evaluation Tool FPGA image, which fills in approximately 2 ms at 1.6~Gsp/s).
A more practical attack setup would incorporate real-time streaming to external storage and offload a portion of the demodulation pipeline to the FPGA, reducing the effective data rate and overall storage requirements.

After samples are received, a band-pass filter is centered around the interrogation frequency with a 125 MHz bandwidth to attenuate ambient noise.
Figure~\ref{fig:PSD} shows the measured spectrum, before and after filtering at an interrogation frequency of 840 MHz.
The target of our measurements is an Ethernet link embedding an HT as described in Section~\ref{sec:trojan_archi} connecting a laptop and a Raspberry Pi 3. 
Samples were collected across multiple carrier frequencies, at output power levels ranging from 0 dBm to 17 dBm, for an IDLE and active Ethernet link.
To evaluate raw attack performance in a controlled environment, all measurements reported here were conducted at a range of 4~m inside an anechoic chamber. 

\section{Data recovery}
\label{sec:data_recovery}

In a conventional Ethernet interface, decoding is straightforward and follows the procedure described in Section~\ref{sec:ethernet_operation}. 
However, due to the errors inherent to the radio interrogation link, directly applying the standard decoding steps would result in excessive bit errors, preventing reliable data recovery.

The decoding algorithms presented below were developed to evaluate the feasibility and effectiveness of information extraction from recorded samples in a realistic interrogation scenario.

Whereas a standard Ethernet controller prioritizes error detection and throughput to maintain a stable communication link, our method focuses on maximizing information retrieval under degraded signal conditions.
To correct errors, we rely heavily on context-dependent decoding, where both past and future data are analyzed to select the most plausible outcome when impossible sequences are detected.

\subsection{Demodulation}
\label{sec:demod}
 
\begin{figure}[!h]
\centering
\includegraphics[width=\columnwidth]{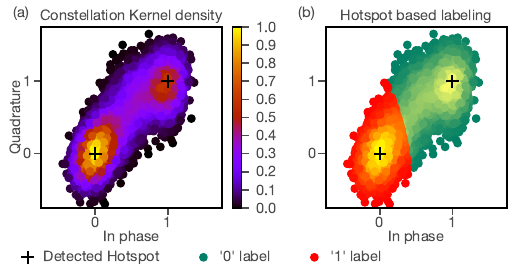}
\caption{(a) Gaussian KDE shown as hue, with highlighted “hotspots” in high-density regions. 
(b) Sample-to-bit attribution using the median of the samples distance to the hotspots as the decision criterion.
Axes are normalized using hotspots positions.}
\label{fig:constel}   
\end{figure}    

After filtering, the recorded samples will comprise the implant response intertwined with the illumination signal. 
The interaction between the illumination signal and the retroreflected states is unpredictable and leads to the data being encoded through both phase and amplitude modulation.
Because extreme states $\Gamma_{\mathrm{on}}$ and $\Gamma_{\mathrm{off}}$ will create sample clusters in the complex plane, we can use the distance between each sample to the densest sample hotspot (i.e., the brightest region in Fig~\ref{fig:constel}(a)) to recover $|D^{+} - D^{-}|$. 
Before proceeding with the different decoding steps, we can verify that this signal is correlated to the original on-wire data.

\begin{figure}[!h]
\centering
\includegraphics[width=\columnwidth]{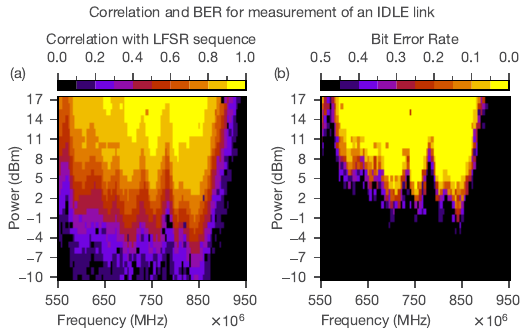}
\caption{(a) Correlation of received backscattered data (IDLE link) with a known 2048-bit scrambler sequence. 
Displayed values correspond to Pearson product–moment correlation coefficients. 
Samples and the scrambler sequence are aligned by maximizing their cross-correlation. 
The scrambler sequence, which is known \textit{a priori}, is resampled and low-pass filtered to match the spectral characteristics of the backscattered signal.  
(b) Bit error rate after LFSR synchronization averaged over $n = 11$ recovered LFSR sequences. }
\label{fig:heatmap_correl_ber}   
\end{figure}

To this end, a simulated scrambler sequence is generated by initializing the LFSR to a random state, and its output is low-pass filtered at 125~MHz to emulate the spectral characteristics of the recovered signal.
Time synchronization between the simulated waveform and the recorded backscattered signal (here known to correspond to an IDLE link) is achieved by maximizing their cross-correlation over one scrambler output sequence period (2047 bits). 
This ensures the recovery of the correct alignment even in the presence of a weak correlation.
The Pearson product–moment correlation coefficient is then computed for all subsequent samples, as shown in Fig~\ref{fig:heatmap_correl_ber}(a).
Correlation values close to zero indicate noise-dominated regions without usable backscattered data, whereas values approaching one denote a strong correspondence between the on-wire data and the recovered backscattered states.

Fig~\ref{fig:heatmap_correl_ber}(a) show a strong correlation for a wide range of frequencies and directly dependent on the power used for illumination, indicating that data recovery with a low error would be achievable. 
  
\subsection{Symbol Attribution}
\label{sec:sample_attribution}

Now that we have confirmed the presence of data correlated with the link activity in the backscattered signal samples, the next step toward data recovery is to extract individual symbols.
 
\paragraph{\textbf{Sample Attribution (hotspot distance)}}
 
Since the implant switches based on $|D^{+} - D^{-}|$, and due to the MLT-3 signaling scheme, symbols associated with $D^{+} \neq D^{-}$ occur twice as often as $D^{+} = D^{-}$.  
These two states produce a biased distribution of samples in the complex plane, as shown in Fig~\ref{fig:constel}(a), where one hotspot appears approximately twice as dense as the other.

Constellation hotspots are identified using Gaussian kernel density estimation (KDE), from which the centers of the two highest-density clusters are extracted. 
Next, the distance from each sample to both hotspots is computed, where a value of 0 indicates perfect alignment with the first hotspot, and 1 with the second (see Fig~\ref{fig:constel}(b)).

\begin{figure}[!h]
\centering
\includegraphics[width=\columnwidth]{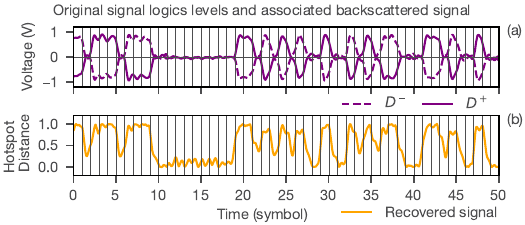}
\caption{(a) MLT-3 logic levels obtained from on-wire captures using an oscilloscope.  
The corresponding backscattered signal recovered by illuminating an implant is presented in (b) (logic levels extracted from sample distances to hotspots). 
Despite the degraded waveform, it remains possible to distinguish between $D^{+} = D^{-}$ and $D^{+} \neq D^{-}$. 
All samples are normalized within [-1,1] for oscilloscope measurements and [1,0] for retroreflected samples. 
The measurements correspond to the same data sequence, matched in post-processing but not acquired simultaneously.}
\label{fig:compared}   
\end{figure}  

To accurately assign each sample to a logical 0 or 1, the signal is first resampled so that the symbol duration corresponds to an integer number of samples. 
This yields a metric based on $|D^{+} - D^{-}|$, as illustrated in Fig~\ref{fig:compared}. This metric is also used as a confidence indicator for each sample’s labeling. 

Finally, the median of all computed distances is used as a threshold to assign each sample a binary label. 
The median is preferred over the mean due to the biased distribution introduced by the implant.
   
\paragraph{\textbf{Sample alignment and symbol extraction}}

To assign a symbol value to a window of $n$ samples, the samples must be properly aligned. Otherwise, a window may span the boundary between two symbols, leading to incorrect assignments.
To find the correct alignment, we test all possible window offsets within one symbol period (i.e., the number of samples per symbol at 125 MBd). 
For each offset, we compute a coherence metric: within each window, samples labeled '1' and '0' are mapped to $+1$ and $-1$, respectively, and their sum taken in absolute value. 
This value is then averaged over many windows. 
The offset that maximizes this average is selected, as it corresponds to the most consistent labeling within windows.
Once the optimal offset is found, each window is assigned a final symbol and confidence level based on the distribution of its samples.

\paragraph{\textbf{Symbol Attribution (SVC)}}    

To obtain an alternative and more tailored symbol attribution we can train a linear classifier based on a C-Support Vector Classification (SVC) model. 
The classifier was made using scikit-learn \cite{ML:scikit}, which relies on libsvm \cite{ML:libsvm} for its SVM implementations.

This approach requires a dataset of sample windows and their associated symbol labels. 
To obtain such a dataset, we first perform symbol attribution and alignment using the hotspot-distance-based method on a known IDLE link. 
During this training phase, the link is assumed to be in the IDLE state, and the scrambler sequence must be successfully recovered (see Section~\ref{sec:LFSR_recover}).
Under these conditions, symbol attribution errors can be easily corrected, as all transmitted unscrambled bits are equal to 1, thereby enabling the establishment of a ground truth for training the SVC classifier.

Each sample within a symbol period is then used as an input feature.
The trained classifier outputs both the predicted symbol value and an associated confidence metric.
Because the model parameters depend on the interrogation conditions (e.g., illumination settings, hardware configuration, and surrounding environment), retraining is required whenever these conditions change significantly.
 
The benefits of this method, including its improved robustness against decoding error, are discussed in Section~\ref{sec:error_correc}.

\subsection{Bit recovery and unscrambling}
\label{sec:LFSR_recover}
   
\paragraph{\textbf{MLT-3 decoding}}

Once the symbols $S$ are recovered, the scrambled bit $c_{n}$ are decoded according to the following rule:

\[
c_{n} =
\begin{cases}
0, & \text{if } S_{n} = S_{n-1} \\
1, & \text{if } S_{n} \neq S_{n-1}
\end{cases}
\]
 
The confidence metric associated with the symbols is not preserved during this conversion but is later exploited in the implementation of error detection and correction (see Section~\ref{sec:error_correc}).

\paragraph{\textbf{LFSR state recovery}}
  
To descramble each bit of the MLT-3 decoded bitstream, we must first determine the associated scrambler output. When the link is in the IDLE state, the scrambler state $K$ can be reconstructed by taking the bitwise complement of 11 consecutive bits (as described in Section~\ref{sec:pld}).

To enhance the robustness of scrambler state recovery against bit misattribution, we extend the standard recovery procedure by using multiple ($N$) 11-bit sequences indexed by $j$. For each 11-bit sequence, the associated LFSR state $K^{j}$ is recovered as:
\[  
K^{j} = \neg c[11j: 11(j+1)[
\] 

To synchronize all $K^{j}$ sequences, each recovered scrambler state is advanced by a number of iterations equal to the full sequence period minus its positional offset, i.e., $2047 - 11j$. This process realigns all states with the first recovered sequence and, in the absence of errors, should yield identical bit sequences.
The set of possible LFSR states ${K^{j}}$ is then averaged and rounded to the nearest integer, yielding the most probable current scrambler state.
 
In our case, the number of sequences chosen was 11 as no significant improvement in BER was observed for higher values of $N$.
The choice of N is dependent on the link activity as a heavily used link may not provide sufficient IDLE time to accommodate higher $N$. 

It should also be noted that using $N = 2$ can increase the error rate, as this accumulates potential errors over 22 bits without benefiting from averaging.
 
To assess the performance of the proposed scrambler recovery method, we compare it to the correlation baseline shown in Fig~\ref{fig:heatmap_correl_ber}(a).
Using the same measurement associated to the IDLE link and applying the Bit Recovery and LFSR State Recovery procedures, we should obtain a sequence comprised only of '1'.
Hence, we can compute the BER as the ratio of decoded '1' to the total number of recovered bits. 
The resulting BER is shown in Fig~\ref{fig:heatmap_correl_ber}(b) and closely follows the trend of the correlation plot in Fig~\ref{fig:heatmap_correl_ber}(a).
Since successful LFSR state recovery is essential for subsequent bit decoding, the BER sharply increases in regions where the scrambler synchronization fails.

\subsection{Bit alignment recovery}

Once the bitstream has been fully decoded, the next step is to reconstruct data bytes using 4B/5B translation. Prior to this step, the correct alignment of the 5-bit codes must be determined.
In a conventional 100BASE-TX link, where decoding errors are infrequent, alignment can be identified by detecting any '0' within an otherwise IDLE bitstream, which indicates the start of a packet (i.e., the J and K symbols in Table~\ref{tab:4B5B_table}).
In this case, however, the presence of decoding errors prevents the use of this method.

Instead, we perform the 4B/5B translation for all five possible bit alignments on a set of non-IDLE bit sequences.
For each alignment, the number of invalid 5-bit codes obtained is counted, and the alignment yielding the fewest invalid codes is selected as the most probable alignment. 
  
\subsection{4B/5B code translation and error correction}
\label{sec:error_correc}

\begin{figure*}[!t]
\centering
\includegraphics[width=\textwidth]{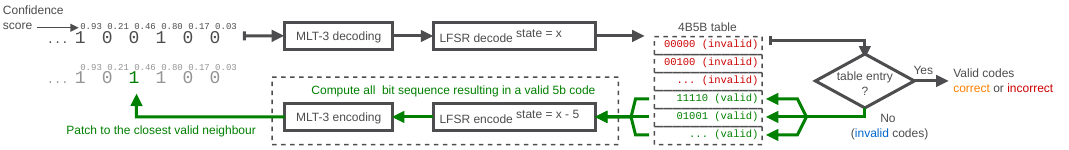}
\caption{Schematic of the error-correction process, in which codes are mapped to their nearest most probable neighbor. Green arrows denote the additional steps compared to conventional decoding.}
\label{fig:patch}  
\end{figure*} 
 
Once the correct alignment has been established, each 5-bit code must be translated into a 4-bit nibble using the mapping in Table~\ref{tab:4B5B_table}. Pairs of nibbles are then grouped into bytes.  
Although this decoding process is straightforward, it propagates errors: a single decoding error within a 6-symbol sequence corrupts the corresponding 5-bit code and its associated 4-bit nibble, thereby altering an entire byte.
 
Because of how a single error propagates, we now aim to correct as many errors as possible.
To devise an error detection and correction mechanism and evaluate its benefit, we will use the following terminology: 

Any 6-symbol sequence results in a decoded code that is either:
\begin{itemize}
  \item \textbf{Uncorrupted}: no symbol is in error.
  \item \textbf{Corrupted}: at least one symbol is erroneous.
\end{itemize}

When translating a 5-bit sequence into a 4-bit nibble using the lookup table, each decoded code can be classified as:
\begin{itemize}
  \item \textbf{Valid}: corresponds to an entry in the translation table.
  \item \textbf{Invalid}: does not correspond to any entry in the table.
\end{itemize}
An invalid code necessarily originates from a corrupted sequence; however, a valid code may still result from a corrupted sequence (i.e., it may map to an incorrect table entry).
 
Valid codes are further classified as:
\begin{itemize}
  \item \textbf{Correct}: matches the originally transmitted data.
  \item \textbf{Incorrect}: differs from the original data.
\end{itemize}
 
Although the 100BASE-TX standard does not include an error detection or correction mechanism, we can exploit the properties of the 4B/5B translation.
Excluding control and reserved symbols, only 20 out of the 32 possible 5-bit combinations are valid (Table~\ref{tab:4B5B_table}).
When packet boundaries are known (which is assumed here), this set is further reduced to 16 valid codes.
Any decoded code outside this subset can therefore be classified as invalid, allowing the detection of errors and an attempt at their correction.

Once errors are detected, we can use the following correction procedure (illustrated in Fig.~\ref{fig:patch}). 
For each invalid code, all candidate 6-symbol sequences producing valid 5-bit codes are enumerated. 
The most likely sequence is selected based on similarity between hypothesized 6-symbol sequences and the associated confidence metrics described in Section~\ref{sec:sample_attribution}. 
If successful, this procedure yields either a valid and correct code or a valid but incorrect code if the correction was unsuccessful.

\subsection{Error correction performance}
\label{sec:error_correc_perf}

\begin{figure*}[!t]
\centering
\includegraphics[width=\textwidth]{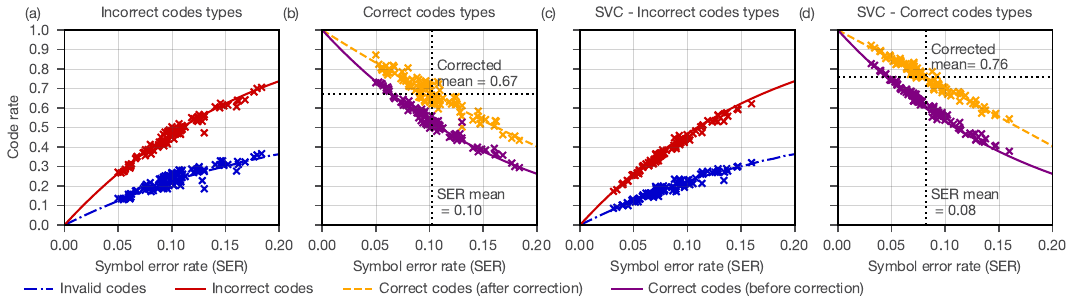}
\caption{(a,c) Proportion of incorrect codes (red), including a subset of invalid codes (blue), as a function of the symbol error rate (SER) of the analyzed data. 
(b,d) Proportion of correct codes, either originally uncorrupted (purple) or obtained after correction (orange). \\
(a,b) SER and correction made from median based discrimination see Sec~\ref{sec:sample_attribution}(a).\\
(c,d) SER and correction made from SVC based discrimination see Sec~\ref{sec:sample_attribution}(b) each point represents the performance of a custom SVC model. \\
The identification of invalid codes is based solely on the data-byte code table; special codes are excluded.}
\label{fig:error_correc}  
\end{figure*} 

To estimate the performance of this algorithm, a subset of the same dataset associated to an IDLE link used in Section~\ref{sec:sample_attribution} is used. 
In this case, all 5-bit codes should be $\mathrm{11111}$, which is not a valid entry in the table used for error detection.
Because of this, and solely for the purpose of evaluating the benefits of this method, a custom 4B/5B table was derived by XORing all 5-bit code entries with $\mathrm{00001}$. This transformation inserts $\mathrm{11111}$ into the valid set while preserving the relative Hamming distances between entries, thus preserving the validity of the evaluation. When decoding data on an active link the original 16 entries of DATA codes should be used. 

To quantify the error rate in the decoded data, we use the code error rate (CER), denoted by $p_{c}$.  
Since each code depends on a MLT-3 encoded 6-symbol sequence, the expected CER is defined as the probability that at least one error occurs within this window (excluding the case where all symbols are inverted):
\[
p_{c} = 1 - (1 - p_{e})^{6} + p_{e}^{6}
\]
where $p_{e}$ denotes the symbol error rate (SER). 
A value of $p_{e} = 0$ corresponds to an error-free sequence, while $p_{e} = 0.5$ yields a fully randomized sequence.
 
In Fig.~\ref{fig:error_correc}(a), a subset of incorrect codes (red) are invalid (blue) and thus do not match any entry in the 4B/5B table; these codes can be targeted for correction.
Prior to correction, only uncorrupted codes (purple) are correctly decoded, as shown in Fig.~\ref{fig:error_correc}(b). 
Applying the proposed correction to invalid codes increases the proportion of correctly decoded codes by approximately $10~\%$ (orange).

A similar evaluation was conducted for the SVC-based bit attribution and confidence estimation approach (see Figure~\ref{fig:error_correc}(c)).
In this case, we observed both a reduction in the average symbol error probability ($\overline{p_{e}} \simeq 0.08$ compared to $0.1$) and an improvement in the code error rate (CER) ($\overline{p_{c}} \simeq 0.24$ compared to $0.33$) see Figure~\ref{fig:error_correc}(d). 
These results indicate both improved initial symbol classification and a more reliable confidence metric, enabling more effective error correction.

\subsection{Packets extraction}

\begin{figure*}[!t]
\centering
\includegraphics[width=\textwidth]{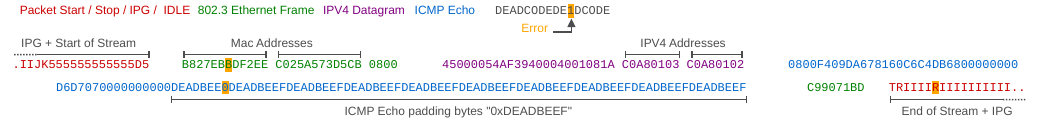}
\caption{Examples of a recovered Ethernet packet obtained from backscattered data after decoding and error correction.
Layer structure and remaining byte-level errors are highlighted.}
\label{fig:packet}  
\end{figure*} 

After applying the previously described error-correcting steps to recordings of an active Ethernet link, we are able to reconstruct packets that are either error-free or exhibit only rare, isolated errors.
To simplify the analysis and ensure a high frequency of short, well-defined frames, the link was flooded with ICMP echo (ping) requests  whose payloads were padded with the sequence $\mathrm{0xDEADBEEF}$, enabling easy identification of transmission errors.

Figure~\ref{fig:packet} shows an example of a successfully recovered packet (BER around 0.01); the main protocol fields remain intact and only a few isolated byte-level errors are present. 
All essential information within the frame (MACs / IPs addresses, payload) remains readable, and the payload pattern is clearly identifiable. 
The same decoding and interpretation procedure generalizes to longer frames and other traffic patterns that contain known structure.

\section{Replication in an office setting and theoretical interrogation range}
\label{sec:realistic}

\begin{figure}[!h]
\centering
\includegraphics[width=\columnwidth]{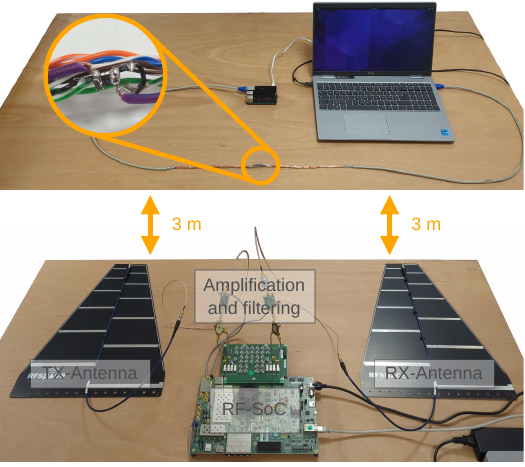}
\caption{Reflecthernet attack setup. (Top) The target, consisting of a laptop connected to a Raspberry Pi through an Ethernet cable embedding the HT (inset). (Bottom) The attack setup, consisting of an RF SoC and two log-periodic antennas.}
\label{fig:interog}   
\end{figure}    

The attack was reproduced outside an anechoic chamber in a standard office environment. 
The experimental setup is shown in Fig.~\ref{fig:interog}, where both the attacker hardware and the target are placed on standard office tables. 
The attack is performed under line-of-sight (LOS) conditions at a distance of 3~m, with an illumination power of 13~dBm.
Under these conditions, a BER below 0.1 was achieved, indicating that the technique remains viable under realistic indoor propagation conditions.
  
While achieving long interrogation ranges would require a setup with higher amplification, we can estimate the required power as a function of distance. 
In free space, and neglecting platform-dependent propagation factors, the received power from a retroreflector implant can be approximated by the radar equation for a backscattering target using the same antenna for transmission and reception:

\[ 
P_\mathrm{r}(\Gamma_{\mathrm{on}},\Gamma_{\mathrm{off}}) =
\frac{P_\mathrm{t}  G^{2}  \lambda^{2} \sigma(\Gamma_{\mathrm{on}},\Gamma_{\mathrm{off}})}
{(4\pi)^{3} R^{4}} 
\] 

where $P_\mathrm{r}$ and $P_\mathrm{t}$ are the received and transmitted powers, respectively, $G$ is the gain of the transmitting and receiving antennas (assumed to be identical), $\lambda$ is the carrier wavelength, $\sigma(\Gamma_{\mathrm{on}},\Gamma_{\mathrm{off}})$ is the implant's radar cross-section which depends on the implant reflection state (here parameterized by the load states $\Gamma_{\mathrm{on}}$ and $\Gamma_{\mathrm{off}}$), and $R$ is the range between the antenna system and the implant. 

The primary factor governing the achievable interrogation distance in RFRA systems is the transmitted (illumination) power. 
Because the incident wave must propagate to the implant and the backscattered signal must then return to the receiving antenna, the received power decreases proportionally to $R^{4}$. 
Consequently, doubling the interrogation range from 3~m to 6~m requires an increase in the transmitted power by a factor of 16 (an increase of approximately 12 dB), corresponding to about 25 dBm for the setup used in this study.   

Along with illumination power, the antenna gain and the implant's design (both its variable load characteristics and its antenna structure) are critical parameters for successful interrogation and data recovery.

Although the transmitted power could, in principle, be further increased to extend the range, practical limitations arise from the receiver's dynamic range. Without appropriate isolation or cancellation mechanisms, the strong illumination signal can saturate the receiver front end, effectively blinding it and preventing demodulation of the much weaker backscattered signal.
  
\section{Conclusion}
\label{sec:conclusion}

We demonstrated the practicality of an RFRA on a 100BASE-TX Ethernet link by recovering complete MAC-layer frames from one traffic direction. Our results extend the effective data rate of RFRA attacks to 125~MBd, demonstrating that retroreflection-based attacks remain feasible even against significantly higher-speed links than those examined in prior works.

The attack was successfully demonstrated at a distance of 3~meters in an office environment. 
An analysis of the BER as a function of the transmitted power in an anechoic chamber suggests that substantially longer interrogation ranges are achievable for an attacker.  

A custom demodulation and interpretation pipeline was created, enabling mitigation of errors introduced by the radio channel and the interrogation process. 
The pipeline exploits the 4B/5B code used in the link to detect and correct a subset of errors; the same methodology could be adapted to other line codes and link technologies targeted by RFRA.

The experiments presented here target a single direction of the full-duplex Ethernet link. 
Extending the attack to both directions would be possible by implanting a second retroreflector HT on the other differential pair and performing a multi-Trojan interception by leveraging frequency diversity, as demonstrated in \cite{11176387}.

\section*{Acknowledgments}

This publication is supported by the European Union through European Regional
Development Fund (ERDF), Ministry of Higher Education and Research, CNRS, Brittany region,
Conseils Départementaux d’Ille-et-Vilaine and Côtes d’Armor, Rennes Métropole, and Lannion
Trégor Communauté, through the CPER Project CyMoCod, by the French government through the ``Agence Nationale de la Recherche" (ANR) both under Grant ANR-22-CPJ1-0070-01 and as part of the ``programme d'Investissements d'Avenir" (PIA) under Grant ANR-18-EURE-0004.

\bibliographystyle{IEEEtran}
\bibliography{IEEEabrv.bib,biblio.bib}

\vfill

\end{document}